\documentclass{article}
\usepackage{nameref}

\usepackage[english]{babel}

\usepackage[psamsfonts]{amsfonts}
\usepackage[dvipsone]{epsfig}
\RequirePackage{graphics,leqno}
\usepackage[LY1]{fontenc}
\usepackage{bm}

\newtheorem{theorem}{Theorem}[section]

\newtheorem{defn}{Definition}[section]

\newtheorem{note}{Remark}[section]

\makeatletter
\newdimen\mathindent
\def\equation{\@beginparpenalty\predisplaypenalty
  \@endparpenalty\postdisplaypenalty
  \refstepcounter{equation}\trivlist \item[]\leavevmode
  \hbox to\linewidth\bgroup $\m@th\displaystyle\hskip\mathindent}

\def\endequation{$\hfil\displaywidth\linewidth\egroup\hskip-12pt\@eqnnum\endtrivlist}
\makeatother

\begin{document}

%
%
%
%
%
%
%
%


\title{A proposal of a faster variant of known provably  secure PRBGs}

\author{A. Corbo Esposito and\newline
F. Didone}


\begin{abstract}
We make a new proposal about how to use in an effective way a CSPRBG (Computationally Secure Pseudo Random Bit Generator) for cryptographic purposes.
We introduce the definitions of TCSPRBG (Typical CSPRBG) and SCSPRBG (Special CSPRBG). In particular the definition of SCSPRBG synthetizes in a simple way our proposal of how to modify a CSPRBG in order to achieve a higher throughput rate, while retaining some essential features of its computational security.

We then summarize which should be, in our opinion, a "standard way" to use a CSPRBG for cryptographic purposes. We eventually present as an application, a particular SCSPRBG for which we can achieve throughput rates greater than $100$ Mbits/sec on current mobile devices. 

\end{abstract}

\maketitle


\section{Introduction}

In this paper we propose an encryption scheme based on a CSPRBG that we believe has some innovative features. 

It is well known that a CSPRBG can be used to communicate encrypted data. In particular the computational work needed to ensure cryptographic security can be carried out outside the temporal interval of encryption-decryption of data: in fact the encryption-decryption phase consists of two xor with a same piece of the string produced by CSPRBG, therefore its execution time is negligible. We describe this aspect saying that the encryption method has an execution time with zero latency, just to highlight that this scheme can be applied to real-time communications. Moreover it is clear that the maximum rate allowed for communication is only bounded by the throughput rate of the CSPRBG. We can even replace the word “maximum” with word “average” in the previous statement if we use some appropriate buffering algorithm.

We first take into account the existence of PRBGs for which security results have been proved. In particular we consider RSAPRBG and QUAD, for which security  results have been proved in papers [RSA] , [QUAD1], [QUAD2]. In the paper [ARTICOLO2] we carried out a Java software implementation of both these PRBGs putting in place all knowledge and tricks we know, obtaining in both cases a throughput rate of several Mbit/sec (see [], pag…); such results seem to be better of the ones forecasted in [], pag… for RSAPRBG and somewhat  not bed for QUAD, giving the results obtained in [], pag… for a hardware implementation of QUAD on a FPGA.

Such throughput rates, nevertheless, in our opinion are insufficient for some kind of applications (namely encryption of audio/video data), since while they could be improved by a hardware implementation, they are actually obtained used with CPU load of 100

In order to obtain a faster variant of these CSPRBGs we introduce the definitions of two particular classes of them, namely the class TCSPRBG that turns out to contain both RSAPRBG and QUAD, and the class SCSPRBG; a PRBG belonging to this class starts from a correspondent TCSPRBG and uses one more one-way function w as a “local” expander of the bit stream, in order to obtain a higher throughput rate while basically maintaining the same security (even if function w could be inverted in some case, this does not compromise the security of the bit stream, see proposition X.X). To give a precise idea of the throughput rates we can obtain, we consider a particular choice for the function w. We call it the “pick” function. Again paper [] provides numerical results of a java software implementation of this function, see….

In the final section we show how the inversion of the “pick” function is the inversion of a (very particular) quadratic system in GF(2). Moreover we also clarify some relations between the “pick” function and the function f used in QUAD.

Then we develop some strategies to invert the “pick” function. We start from the simplest ones or from some adaptation of known strategies for inversion of quadratic systems to our particular case, trying our best to exploit particularities of the “pick” function to make them faster. The best time estimated for the inversion “pick” function seem to confirm that such inversion is practically not possible for the choice of parameters we have chosen in [].

We anyway aren’t able to obtain sharp lower bounds for the time needed for inversion.

It seems to us however that the combined effect of proposition X.X and the estimated times for the inversion of the “pick” function is that an SCSPRBG based on a correspondent TCSPRBG given by RSAPRBG or QUAD coupled with the choice of “pick” function as w deserves some attention.

{\it Organization.} Section 2 contains notations and definitions. In the section 3 we introduce the definitions of TCSPRBG and SCSPRBG along with some comments and remarks. In the section 4 we describe the method; in the section 5 we discuss the applications of the method on existing mobile devices.


\section{Preliminary notations and definitions}

{\it Notation.} We use $i\oplus j$ to denote the bitwise xor of integers $i$ and $j$ and $||$ to denote the concatenation of two sequences.

\subsection{One way function}

There are some types of functions that play a significant roles in cryptography. One of these types is the one way function. We adopt for our purposes the following definitions of one-way function from \cite{Goldreich}.

\begin{defn}[Strong one way function]\label{OWFs}
A function $f:\{0,1\}^n\to \{0,1\}^m$ with $m=O(n)$ is called {\it (strongly) one way function} if the following two condition hold:
\begin{enumerate}
\item Easy to compute: There exists a deterministic polynomial time algorithm $A$ such that on input $x$ algorithm $A$ outputs $f(x)$ (i.e. $A(x)=f(x)$).
\item Hard to invert: For every probabilistic polynomial time algorithm $A'$, every positive polynomial $p(\cdot )$ and all sufficiently large $n'$s,
$$Pr[A'(f(U_n))\in f^{-1}(f(U_n))]<\frac{1}{p(n)}$$
\end{enumerate}
where $U_n$ denotes a random variable uniformly distributed over $\{0,1\}^n$.
\end{defn}

\begin{defn}[Weak one way function]
A function $f:\{0,1\}^n\to \{0,1\}^m$ with $m=O(n)$ is called {\it (weak) one way function} if the following two condition hold:
\begin{enumerate}
\item Easy to compute: There exists a deterministic polynomial time algorithm $A$ such that on input $X$ algorithm $A$ outputs $f(x)$ (i.e. $A(x)=f(x)$).
\item Slightly hard to invert: There exists a polynomial $p(\cdot )$ such that for every probabilistic polynomial time algorithm $A'$ and all sufficiently large $n'$s:
$$Pr[A'(f(U_n))\not\in f^{-1}(f(U_n))]>\frac{1}{p(n)}$$
\end{enumerate}
\end{defn}

\subsection[Random bit generator]{Random bit generator}

\begin{defn}
A {\it random bit generator} is an algorithm with outputs a sequence of statistically independent and unbiased binary digits.
\end{defn}

\begin{note}
It is possible to use a random bit generator to generate random numbers. For example a integer number in the interval $[0,n]$ can be obtained by generating a random bit sequence of length $\lfloor log_2 n\rfloor+1$ and converting it to an integer.
\end{note}

A true random bit generator requires a naturally occurring source of randomness. It is difficult to design  a hardware device or software program that produces uncorrelated bit sequences exploiting this randomness. Moreover this kind of generator is influenced by external features so it must be periodically tested.

There are two kind of true random bit generator:

\begin{itemize}
\item[-] hardware-based generator (that exploits the randomness which occurs in some physical phenomena);
\item[-] software-based generator (that can be based on the system clock or on the content of input/output buffers, and this is more difficult to design than previous).
\end{itemize}

\subsection[Pseudo random bit generator]{Pseudo random bit generator}

\begin{defn}\label{PRBG}
A {\it pseudo-random bit generator (PRBG)} is a deterministic algorithm which, given a truly-random binary sequence of length $k$,  outputs a binary  sequence of length $l>>k$ which appears to be random. The input to the PRBG is called {\it seed} and the output is called a {\it pseudo-random bit sequence}. 
\end{defn}

The output to the PRBG is not random because the number of output sequences is a small fraction, $2^k/2^l$ of all binary sequences of length $l$.

The idea is to consider a small truly-random sequence and to expand it in a longer sequence. In this way a possible adversary can not easily distinguish between output sequences of the PRBG and truly-random sequences of length $l$. Therefore there are statistical tests that represent necessary conditions but not sufficient so that a generator may be secure. In fact it is impossible to prove mathematically that the output of a generator is truly-random. Tests are probabilistic.

A minimum security condition for a random sequence generator is that the length $k$ of the random seed should be sufficiently large so that a search over $2^k$ elements is infeasible for the adversary.

The seed must have two properties:

\begin{itemize}
\item[-] the output sequence of PRGB must be statistically indistinguishable from the truly-random sequence;
\item[-] output bits must be unpredictable by the adversary that has limited computational resources.
\end{itemize}

We now have to specify what "appears to be random" main.

\begin{defn}
A PRGB is said to pass all polynomial-time statistical tests, and therefore can be considered as a cryptographic secure PRGB, if no polynomial-time algorithm can distinguish between an output sequence of the generator and truly-random sequence with probability significantly greater than $1/2$.
\end{defn}

\begin{defn}
A pseudo-random bit generator is said to pass the next-bit test if there is no polynomial-time algorithm which, on input of the first $l$ bits of an output sequence $s$, can predict the $(l+1)^{st}$ bit of $s$ with probability significantly greater than $\frac{1}{2}$.
\end{defn}

\begin{note}
A pseudo-random bit generator passes the next-bit test if and only if it passes all polynomial-time statistical tests.
\end{note}

\begin{defn}\label{CSPRGB}
A PRBG that passes the next-bit test (possibly under some plausible but unproved mathematical assumption such as the intractability of factoring integers) is called a {\it cryptographically secure pseudo-random bit generator (CSPRBG)}.
\end{defn}


\section[Typical CSPRBG]{Typical CSPRBG}

\begin{defn}\label{TCSPRBG}
A {\it typical-CSPRBG} (denoted with TCSPRBG) is defined as follows. Computed the initial value $x_0$ from a true random seed $u_0$  after an initialization phase, let $f$ be an one-way invertible function and $g$ be an one-way not (invertible) function. Compute:
$$x_{i+1}=f(p_1,...,p_k,x_i)\quad \forall i=0,1,...$$
where $p_1,...,p_k$ are fixed and known parameters,
$$y_{i}=g(x_i)\quad \forall i=1,2,...$$
The output sequence to pseudo-random generator is:
$$s=y_{1}||y_{2}||...$$
where $||$ denotes the  concatenation.
\end{defn}

The output sequence depend on the properties of the one-way function used therefore it may be necessary $y_i$ only keeps some bits of the output values $x_i$ in order to remove possible correlation between successive values. Therefore the  function $g$ is typically a projection.

\begin{defn}\label{TCSPRBGproblem}
The {\it TCSPRBG inversion  problem}  is the following: given $y_1||y_2||...||y_n$, find $x_{n+1}$.
\end{defn}

\begin{defn}
The {\it TCSPRBG partial inversion  problem}  is the following: given $y_1||y_2||...||y_n$, find $y_{n+1}$.
\end{defn}

\begin{note}
The idea behind previous definition  is the following: 
\begin{itemize}
\item[-] if we adversaly solve the inversion problem defined by \ref{TCSPRBGproblem} we clearly can predicted all the future output of the PRBG, and the security in this case is completaly broken;
\item[-] if we adversaly can solve for some values of $n$  (but not for every $n$) the partial inversion problem it can still predicted part of the future output of the PRBG and security partial broken.
\end{itemize}
\end{note}

\begin{note}
The "CS" in  "TCSPRBG" stands for cryptographically secure. There can not be an unconditioned proof of this feature (the same is true for the existence of a one-way function, see \cite{Goldreich}). All existing security proof are conditioned  to conjectures of the following types:
\begin{itemize}
\item[-] $P\neq NP$;
\item[-] on unsolvability of well known problems in less than a certain certain time.
\end{itemize}
\end{note}
Two examples of TCSPRBG we consider in the present thesis are:
\begin{itemize}
\item[-] a variant of the Micali Schnorr RSA PRG studied by Steinfield, Pieprzyk and Wang in \cite{ST};
\item[-] QUAD \cite{QUAD},\cite{QUAD2}.
\end{itemize}

The strategy that is followed  to prove that such PRBGs are computational secure is the following:
\begin{itemize}
\item[-] a necessary premise to solve any inversion problem is that sequence $y_1||y_2..||y_n$ is distinguishable from a random sequence;
\item[-] if such sequence (for $n$ not too large) is distinguishable from a random sequence then some well known hard problem could be solved more efficently than it is actually known.
\end{itemize}

In order to give an idea of how secure these CSPRBG are, we report the statement of the main security results obtained by these authors respectively for the RSA PRBG and for the QUAD:

\begin{theorem}\label{teST}
For all $n\geq2^9$, any $(T,\delta)$ distinguisher $D$ for $(n, e, r,l)$-RSAPRG can be converted into a $(T_{INV},\epsilon_{INV})$ inversion algorithm $A$ for the $(n, e, r,w)$-CopRSA problem (with $w=3\log(2l/\delta)+5$) with:
\begin{equation}\label{TS}
T_{INV}=C_S(T+O(l/r\log(e)n^2))
\end{equation}
where:
$$C_S=64 (l/\delta)^2n\log(n)$$
and:
$$\epsilon_{INV}=\delta/9-4/2^{n/2}$$
\end{theorem}

\begin{theorem}\label{TeQUAD}
Let $L=\lambda (k-1)n$ be the number of key-stream bits produced by in time $\lambda T_S$ using $\lambda$ iterations of our construction. Suppose there exists an algorithm $A$ that distinguishes the $L$-bit key-stream sequence associated with a known randomly chosen system $S$ and an unknown randomly chosen initial internal state $x\in \{0,1\}^n$ from a random $L$-bit sequence in time $T$ with advantage $\epsilon$. Then there exists an algorithm $C$, which given the image $S(x)$ of a randomly chosen (unknown) $n$-bit value $x$ by a randomly chosen $n$-bit to $m$-bit quadratic system $S$ produces a preimage of $S(x)$ with probability at least $\frac{\epsilon}{2^3\lambda}$ over all possible values of $x$ and $S$ in time upper bounded by $T'$.
$$T'=\frac{2^7 n^2 \lambda^2}{\epsilon^2}\bigg(T+(\lambda+2)T_S+\log\bigg(\frac{2^7n\lambda^2}{\epsilon^2}\bigg)\bigg)+\frac{2^7n \lambda^2}{\epsilon^2}T_S$$
\end{theorem}

A very rough interpretation of these theorems can be given us follows:
\begin{itemize}
\item[-] theorem \ref{teST} states that if the sequence produced by RSAPRBG is distinguishable (in a certain time) from a random sequence then an "RSA type" problem (the precise definition of $(n,e,r,w)$-CopRSA is given in definition 4.2 of \cite{ST}) could be solved in less time than the best known attack, i.e. the Coppersmith attack. For the choice $n=6144$ this implies that an output up to $2^{32}$ bits should be secure given the actual computer tecnology ($2^{70}$ instruction needed), see table 1 of \cite{ST};
\item[-] theorem \ref{teQUAD} states that in the sequence produced by QUAD is distinguishable from a random sequence, than a quadratic system in GF2 should be solved in less time than expected by the best known algorithms based on Groebner bases, resulting contradiction ($n>350$).
\end{itemize}

One point in favour of RSAPRBG is that the practical implementation of QUAD is proposed for $n=160$, while one point in favour of QUAD is that the solution of quadratic systems in GF2 is known to be NP-complete.

With regard to the applications of this type of generator it is necessary to take into account the following features:
\begin{enumerate}
\item conjectures that are assumed to obtain security results (e.g. intractability of integer factorization can no longer be true if quantum computers will be realized and introduced in the market);
\item parameters for which the security of these families of generators is guaranteed;
\item the maximum bit-length of output sequence that can be made public so that the system continues to be safe;
\item system throughput. In practice this assessment can not be purely theoretical since it is impossible to anticipate all the operations performed by a generic machine (data movement operations, memory management operations,...).
\end{enumerate}

\subsection[Special TCSPRBG]{Special TCSPRBG}

Now we consider a modification of TCSPRBG. We call it Special TCSPRBG (SCSPRBG for short). 
\begin{defn}\label{defSCSPRBG}
A SCSPRBG is defined as follows. Let $f$ and $g$ be two function as in the definition \ref{TCSPRBG}. Let:
$$x_{i+1}=f(p_1,...,p_k,x_i)$$
since the parameters $p_1,...,p_k$ are fixed and known usually we write $f(x_i)$.
$$y_{i}=g(x_i)$$
$$z_{i}=w(y_i)$$

The function $w$ is a weakly one way function such that:
\begin{itemize}
\item[-] $\#_{bit (z_i)}>>\#_{bit}(y_i)$
\end{itemize}
where $\#_{bit}(\cdot)$ denotes the number of bits of the sequence. 
The output sequence of generator is $z_1||z_2||...$. 
\end{defn}

\begin{defn}\label{correspondent}
The TCSPRBG defined above is set correspondent to the SCSPRBG that uses the same function $f$ and the same function $g$.

We say that a  TCSPRBG and a  SCSPRBG are correspondent if they use the same function $f$ and the same function $g$.
\end{defn}

\begin{defn}\label{IPSCS}
The {\it SCSPRBG inversion problem} is defined as follows: given $z_1||z_2||...||z_n$ find $x_{n+1}$.
\end{defn}

\begin{defn}
The {\it SCSPRBG partial inversion problem} is defined as follows: given $z_1||z_2||...||z_n$ find $y_{n+1}$.
\end{defn}

\begin{note}
The SCSPRBG inversion problems are obviously not easier than the inversion problems for the correpondent TCSPRBG.
\end{note}

\begin{defn}\label{PIPSCS}
The {\it SCSPRBG sub-partial inversion problem} is defined as the following: given $z_1||z_2||...||z_n$ and a subset of the bits of $z_{n+1}$, find $z_{n+1}$.
\end{defn}

\begin{note}
This problem can be much easier than the previous if $w$ is not carefully chosen. To understand this, since the number of bits of $z(n+1)$ is larger than the number of bits of $y(n+1)$, a partial knowledge of $z(n+1)$ could be sufficient to determine $y(n+1)$ and to compute $z(n+1)=w(y(n+1))$. Moreover the previous definition emphasizes that if one adversary can solve the sub-partial inversion problem he actually gains some information.
\end{note}

\begin{note} 
On the other side if one adversary is not able to solve the inversion and partial inversion problem for the SCSPRBG it is reasonable to assume that the sub-partial inversion problem is not easier than the inversion problem for the function $w$ (i.e. given $z(n+1)$, find $y(n+1)$). Therefore if we build an SCSPRBG correspondent to a TCSPRBG (for example correspondent to RSAPRBG or QUAD) we only study the difficulty of the inversion of function $w$.
\end{note}

\begin{note}
The throughput rate of a SCSPRBG can be much higher than the throughput of correspondent TCSPRBG, due to use of the weakly one way function $w$.
\end{note}

Next section will be to devoted to find good candidate for the choice of function $w$.


\section[The function pick $(k,m)$]{The function pick $(k,m)$}

\begin{defn}
Let $k,m$ be numerical and let $l=m\cdot 2^{k-1}$. Let now $M$ be a public known matrix of random bits with $l$ rows and $2l$ (i.e. $m\cdot 2^k$) columns. We call pick $(k,m)$ the following function that transforms a string of $mk$ bits in a string of $l$ bits. The input of $mk$ bits is divided in $ m$ segments of $k$ bits. The $ m\cdot 2^k$ columns of $M$ are arranged in $m$ blocks of $2^k$ columns cach. Each segment of $k$ bits is used to choose ("pick") column among the $2^k$ columns in the each block. 

The output string is the xor of the choosen columns (and for each block). 
\end{defn}

\begin{note}
The ratio between the length of the output string and the input one is obviously $2^{k-1}/k$.
\end{note}

\begin{note}
Let assume that we have a TCSPRBG that, at each iteration outputs $m\cdot k$ bits. Then, if the pick$(k,m)$ function turns out to be a weakly one way function than it can be used as the function $w$ in the definition \ref{defSCSPRBG}.
\end{note}

%

\begin{defn}
The {\it pick(k,m)  inversion problem} is defined as follows: given the matrix $M$ and the output string find the input string.
\end{defn}


We give some estimate of the difficulties of this inversion problem. We fisrt show that to solve the inversion problem is equivalent to solve a system with $m \cdot 2k $ unknowns, $l+m$  linear equations and $m\cdot \bigg( \begin{matrix}2^k // 2\end{matrix}\bigg)$.


Let $a_{h,i,j}$ the element of the Matrix $M$ of row $1<h<l$ and column $0<j<2^k-1$ in the block $1<i<m$. We denote with $x=(j_1,...,j_m)\in (\mathbb{Z_{2^k}})^m$ the input string and $y=(y_1,...,y_l)\in (\mathbb{Z_{2^k}})^m$ the output string. There exist $m$ linear relationship:

\begin{equation}\label{eqLin}
\sum_{i=1}^{m}\sum_{j=0}^{2^k-1}a_{h,i,j}\cdot q_{i,j}=y_h
\end{equation}


with $1\leq h\leq l$ and where $q_{i,j}$ are $m2^k$ variables in $\mathbb{Z}_{/2}$. We have for all $1\leq i\leq m$:

$$ q_{i,j}=
\begin{matrix}
0 \ j\neq j_i\\
1 \  j=j_i
\end{matrix} $$


Therefore $\forall 1\leq i\leq m$ an $q_{i,j}$ is equal to 1.


This corresponds to accompany the system of linear equations \ ref {eqLin} with the following quadratic equations:

\begin{equation}
\begin{matrix}
q_{i,j}q_{i,l}=0\ 0\leq j \leq l\leq 2^k-1\\
\sum_{j=0}^{2^k-1}q_{i,j}=1\ 1\leq i\leq m
\end{matrix}
\end{equation}



We see that an iteration of the QUAD can be traced back to a case in a standard way (much) of a particular iteration of the problem pick. For simplicity we consider only iterations QUAD doubling the number of bits (i.e.  $ k=2 $).  In the QUAD iteration we have $n$ bits and a matrix  with $2n$ rows and $1+n+\bigg( \begin{matrix}n\\ 2 \end{matrix}$ columns. In this case we call $x=(x_1,...,x_n)\in (\mathbb{Z}_{/2})^n$ the input string and $y=(y_1,...,y_{2^n})\in (\mathbb{Z}_{/2})^n$ the output string; the relationship between the two strings are given by the $2n$ equazioni quadratiche:

$$\sum_{1\leq i<j \leq n}a_{i,j}^hx_ix_j+\sum_{i=1}^{n}b_i^hx_i=y_h$$


with $a_{i,j} \in A$, $b_i\in B$, where  $A$ is the matrix with $2n$ rows and $\bigg(\begin{matrix}n\\2\end{matrix}\bigg)$ and $B$ a matrix with $2n$ rows and $n$ columns (for simplicity we don't consider the columns  $c_i^h$). We suppose that $n$ is divisible by $2k$ (where $k$ is the variable in the pick definition). Let:

$$m=\frac{n}{2k}+\frac{2n}{2k}(\frac{n}{2k}-1)=\frac{n(n-k)}{2k^2}$$

%

We divide the index $1,..,n$ in $\frac{n}{2k}$ blocks  $\sigma_1,...,\sigma_{\frac{n}{2k}}$ di $2k$ indeces. Every block is divided in two sub-blocks with the same number of indeces such that $\sigma_i=\tau_i \cup \phi_i$.

$$\sum_{r=1}^{n/(2k)}\bigg( \sum_{i\in \sigma_r} b_i^hx_i + \sum_{i<j con i,j\sigma_r} a_{i,j}^hx_ix_j\bigg)+$$

$$+\sum_{1\leq r<s\leq k}^{n/(2k)}\bigg( \sum_{i\in \sigma_r}\sum_{j\in \sigma_s}  a_{i,j}^hx_ix_j\bigg)+$$

$$+\sum_{1\leq r<s\leq k}^{n/(2k)}\bigg( \sum_{i\in \sigma_r}\sum_{j\in \tau_s}  a_{i,j}^hx_ix_j\bigg)+$$

$$+\sum_{1\leq r<s\leq k}^{n/(2k)}\bigg( \sum_{i\in \tau_r}\sum_{j\in \sigma_s}  a_{i,j}^hx_ix_j\bigg)+$$

$$+\sum_{1\leq r<s\leq k}^{n/(2k)}\bigg( \sum_{i\in \tau_r}\sum_{j\in \tau_s}  a_{i,j}^hx_ix_j\bigg)+$$


We fix $r$, the result of sum ( or xor) depends by only values of $\{ (x_i),i\in \sigma_r\}\in (\mathbb{Z}_{/2})^{2k}$. therefore we can rewrite:

$$ \sum_{r=1}^{n/(2k)}\sum_{j=1}^{2^k}a_{h,r,j}\cdot q_{r,j}$$


where $a_{h,r,j}$ are appropriate values and for all $1\leq r\leq \frac{n}{2k}$ one of $q_{r,j}$ is equal 1.


Likewise each of four sum are rewrite as follows:

$$\sum_{r=\frac{n}{2k}+1}^{\rho_1+\frac{n}{2k}}a_{h,r,j}q_{r,j}...\sum_{r=\rho_3\frac{n}{2k}+1}^{\rho_4+\frac{n}{2k}}$$

%

where  $\rho_{s+1}-\rho_{s}=\frac{n}{2k}\bigg( \frac{n}{2k}-1\bigg)$ where $\rho_0=0$  e $s=1,2,3,4$ and where for all values of $r$ exists a only  $j=j(r)$ tale che $q_{r,j}$.

\section[Efficient encryption method based on a  SCSPRBG]{Efficient encryption method based on a  SCSPRBG}

In this section we present how to use a TCSPRBG or a SCSPRBG encrypt communications.

Suppose two users, Alice and Bob, want to communicate securely. The communication between two users can be of two types:
\begin{itemize}
\item[-] one-way communication
\item[-] two-way communication
\end{itemize}

In an one way communication only one of the two users can send the message and the other can only receive it, while in a two-way communication each user can both transmit and receive messages.

Moreover a two way-communication  can be:
\begin{itemize}
\item[-] symmetrical (it is a communication system in which the speed or quantity of data is the same in both directions, averaged over time, i.e. telephone);
\item[-] asymmetrical (it is a communication system in which the data speed or quantity differs in one direction as compared with the other direction, averaged over time, i.e. ADSL).
\end{itemize}

The encryption method works in both cases in almost the same way.

For simplicity's sake we consider an one-way communication: suppose Alice want to send a message $M$ to Bob. Both use the same TCSPRBG (or SCSPRBG) characterize by a private seed, $x_0$, and by some known parameters, $p_1,...p_k$.

\begin{note}
In the case of two-way communication, the two users must generate two sequence TCSPRBG (or SCSPRBG), one used to send a message from Alice to Bob, the other used to send a message from Bob to Alice.
\end{note}

Then first two users exchange the seed securely (in fact it must be known to only two users). Now Alice and Bob parallel generate a long sequence TCSPRBG (or SCSPRBG), denoted by $S$. The $S$ sequence is managed in FIFO mode (first in, first out), i.e. the first bits products are used to encrypt messages while the new bits products are positioned in the line. 


They use pieces of $S$ sequence, denoted with $PRB_i$,  as keys for symmetric key encryption algorithm which will protect the communication (note that a symmetric encryption algorithm is computationally more advantageous than a public key  encryption algorithm).

Suppose Alice wishs to send to Bob the message $M$ securely.  The message $M$ is divided into $k$ sequences:
$$M=M_1||M_2||\cdots||M_k$$
each sequence $M_i$ is encrypted with a piece $PRB_i$ of the sequence $S$ (having the same bit-length of $M_i$) through a xor operation:
$$ALICE:\ enc(M_i)=M_i\oplus PRB_i\ \forall  i=1,..., k$$
Alice sends the encrypted message to BOB. He is able to decrypt it through another simple xor operation because he already has the sequence $PRB_i$:
$$BOB:\ dec(enc(M_i))=enc(M_i)\oplus PRB_i=M_i\ \forall  i=1,..., k$$

Encryption and decryption time are negligible. Moreover the parallel computation of the chunk $PRB_i$ that we used to xor the message $M_i$ can be done {\it before} its use.

\begin{note}
This scheme amounts to consider the output sequence of a TCSPRBG (SCSPRBG) as a key-stream for a stream cypher. 
\end{note}

\begin{note}
No padding schemes are required for the message $M$ and such type of attacks can not be constructed.
\end{note}
\begin{note}
A possible drawback of the scheme is that the actual bit production rate can be very different for Alice and Bob.
\end{note}

\begin{note}
The scheme works until the rate with the bits are produced and added to the sequence $s(x_0)$ is greater than the rate at which the bits are taken to encrypt (decrypt) messages.
\end{note}

This method is secure if:
\begin{itemize}
\item[-] the key $PRB_i$ is used only once;
\item[-] not a single value $x_i$ can be fully recovered by an attacker;
\item[-] the sequence $S$ is a PRBG.
\end{itemize}

Therefore it is necessary that Alice and BOB communicate through a $CSPRBG$.

Two users must share a common seed $x_0$ to generate the pseudo-random sequence $S$ both for a $TCSPRBG$ that for a $SCSPRBG$. There are two problems:
\begin{enumerate}
\item to exchange the seed $x_0$ securely;
\item to preserve the seed $x_0$ securely until it is used.
\end{enumerate}

A possible solution for the first problem is that users exchange the seed personally. Another solution is to use a public encryption algorithm (as RSA) to exchange the seed. If the seed is exchanged through RSA algorithm it is secure if:
\begin{itemize}  
\item[-] quantum computers not exist;
\item[-] authentication is made.
\end{itemize}

\begin{note}
If parts of the seed are encrypted via RSA algorithm and sent in a cross-way between Alice and Bob we can observe that attacker should break both RSA-keys of Alice and Bob.
\end{note}

\begin{note}
In the present work we don't consider in detail the following problems:
\begin{itemize}
\item[-] how to produce and share truly random seed between Alice and Bob;
\item[-] authentication problems (Man in the middle);
\item[-] moreover we only generically address the problem of secret data protection for Alice and Bob. 
\end{itemize}
\end{note}

\begin{note}
We prefer to use RSA algorithm rather than elliptic curves for various practical reasons:
\begin{itemize}
\item[-] for simplicity's sake;
\item[-] RSA cryptanalysis has been most studied than that of elliptic curves.
\item[-] we believe that the actual implementation of RSA, given its simplicity, can be less prone to fatal mistakes.
\end{itemize}

\end{note}

The second problem can be solved generating and exchanging the seed shortly before its use. In this
case all stages of generation of the seed must be protected too.

\begin{note}
In this work we will not consider in details these two problems or the problem to generate really random sequences. We will only focus on the use and efficiency of a TCSPRBG and a SCSPRBG.
\end{note}

\section{An example of an efficient SCSPRBG}

In this section we want to give an example of an SCSPRBG for which can be given some security results. Since the inversion problem (see def.\ref{IPSCS}) for a SCSPRBG is not easier than the inversion problem for the correspondent TCSPRBG, the security results that are valid for the correspondent TCSPRBG still hold.

The hardness of SCSPRBG partial inversion problem (see def. \ref{PIPSCS}) is directly connected with the choice of the function $w$. We can think of many examples to construct the $w$ function as we require that it is only a weakly one-way function. We outline one example that was inspired to us by QUAD.

The crypto-analysis of such example is far from complete, but through it we want to illustrate a basic rule, in our opinion,  to choose the $w$ function, i.e. the simplicity; in fact if the function structure is relatively simple its safety will be easier to prove. Furthermore in the opposite case it will be easier to build  an attack that proves its insecurity, thus prompting the need to change the choice of the function.

\subsection[Possible $w$ function]{Possible $w$ function}

We suppose to have a system of $n$ linear equations in $2n$ unknowns in $GF(2)$, with $n$ sufficiently large (each coefficient is represented by a bit).

We suppose:
$$2n=2^l=2^l2^{l-h}$$
with $h<l$. $a_{ij}$ denotes the $j$-th coefficient of $i$-th equation; $A^j$ denotes the $j$-th column of the matrix $A$.

We split up the column in $2^{l-h}$ blocks, each containing $2^h$ columns. Now we choose one column for each block and compute the xor of the selected columns:
\begin{equation}\label{xorA}
A^{i_1}\oplus...\oplus A^{i_s}
\end{equation}
where $s=2^{l-h}$ and $i_j=(j-1)2^h+1+r_j$ where $r_j$ is an integer such that $0\leq r_j\leq 2^{h-1}$.

Then to make that choice, we must specify $r_j$ for $1\leq j\leq 2^{l-h}$; each $r_j$ consists of $h$ bits, totalling $h2^{l-h}$ bits. This sequence represents the input of the $w$ function and the result of the operation \ref{xorA} represents the output.

Before making an assessment of the safety of such system, we consider its computational efficiency. The system requires $\frac{2^{l-h}-1}{d}$ xor/bit where $d$ is the number of bits of a word (i.e. $d=32$ or $d=64$). The multiplication factor $R$ is the ratio between the bit length of output sequence and the bit length of the input sequence:
$$R=\frac{2^{l-1}}{h2^{l-h}}=\frac{2^{h-1}}{h}$$

\begin{note}\label{r1}
We consider a numerical example. if $2n=8192$ and $h=6$ we need only 4 xor/bit for $d=32$, and we obtain $R=32/6$. 

The computation of $w$ is then at least 25 times faster than QUAD (see equation \ref{xorMQ}). To compute the speed of the SCSPRBG we have to take into a count the time spent for the iteration of $f$ (the time spent in the computation of $g$ is negligible). 

We suppose that an iteration of CSPRBG (considering only the function $f$ and $g$) costs $T_1$ cycles/bit and that the function $w$ costs $T_2$ cycles/bit with $R$ as multiplication factor. Each bit of first iteration generates $R$ bit used for $w$ function; therefore the cost is:
\begin{equation}\label{cost}
\bigg(\frac{T_1}{R}+T_2\bigg)\ cycles/bit
\end{equation}
In the case of RSAPRG authors declare a value $T_1\sim 3000$ cycles/bit (see \cite{ST}) while in the case of QUAD it is reasonable to estimate 200-400 cycles/bit. Therefore even if we use QUAD as correspondent TCSPRBG we can neglect $T_2$ with respect to $T_1/R$ and we can conclude that the speed of our system is roughly 5 times greater than QUAD.
\end{note}

\begin{note}\label{r2}
We want to compare the amount of memory required by the system described above rather to a MQ system. For instance for $n=8192$, a system of this type requires a memory of $\sim 4$ Mbytes. We remember that a MQ system with 320 equations in 160 unknowns requires $1/2$ Mbytes of memory. Therefore in our case we need a storage space a 8 times bigger than a MQ system.
\end{note}

\begin{note}
We can iterate the function $w$ a few times. This can be done in order to get a higher speed, even if $j$ is rather low, e.g. $j=2$ or $j=3$, we can achieve a much higher speed than the correspondent TCSPRBG. Such speed can be computed in a similar way to \ref{cost}, but in any case will obviously be bounded by $T_2$ cycles/bit.
\end{note}

\subsection{Crypto-analysis considerations}

We have just started the study of the proposed system above, but we want to mention some considerations for completeness.

We analyzed is how difficult it is to invert $w$; more precisely the problem is the following: 
\begin{itemize}
\item[-] given $A^{i_1}\oplus ...\oplus A^{i_s}$, find the sequence $r_1||...||r_s$.
\end{itemize}
This problem does not correspond to the definition of partial inversion problem we give (see def. \ref{PIPSCS}) but it seems to us that its computational complexity is a good indicator for the computational complexity of the partial inversion problem.

We observe that the choice of exactly one column of each block of length $2^h$ corresponds to the addition of the following equation for each block:
\begin{equation}\label{i}
\sum_{i\in block}x_i=1
\end{equation}
\begin{equation}\label{ii}
x_ix_j=0\quad \forall i,j\in block
\end{equation}

The equations \ref{i} are $2^{l-h}$ linear equations (one for each block) and \ref{ii} are $2^{l-1}(2^h-1)$ quadratic equations. We observe that the equations \ref{ii} are not independent. Equations (\ref{ii}) can be replaced by the following minimal set of quadratic equations for each block:
\begin{equation}\label{iii}
x_ix_j=0\quad \forall i,j\in block:\ i+j=0\ in\ GF(2)
\end{equation}
Equations \ref{iii} give that in a single block there is at most one $x_i$ equal to 1 when the index $i$ is odd and at most one $x_i$ equal to 1 when the index is even. Then the equation \ref{i} gives us that exactly one $x_i$ is equal to 1 for each block.  The total number of equations \ref{iii} is $2^{l-1}(2^{h-1}-1)$. Therefore we get an MQ system with this particular structure:
\begin{itemize}
\item[-]$2^l$ unknowns;
\item[-] $2^{l-1}+2^{l-h}$ linear equations;
\item[-] $2^{l-1}(2^{h-1}-1)$ quadratic equations.
\end{itemize}
The best way to reduce the complexity of the system seems to us to use Gauss elimination to possible eliminate $2^{l-1}+2^{l-h}$ unknowns we will then be left with a system with the following structure:
\begin{itemize}
\item[-] $2^{l-1}-2^{l-h}$ unknowns;
\item[-] $2^{l-1}(2^{h-1}-1)$ equations of second degree.
\end{itemize}

\subsection{Three strategies to find the solution of the system}

Now we want to illustrate three strategies to find the solution of our system. We assume that we are left with the first $2^{l-1}-2^{l-h}$ unknowns.

\subsection[First strategy]{First strategy}

 A straightforward one  is the following:
\begin{enumerate}
\item For each block of left unknowns we have to choice one variable to be put equal to one. This gives $2^{h(2^{l-h-1}-2^{l-h})}$ possibilities.
\item To compute the other variables and verify if they too respect the equations \ref{iii}; in this case (very probably) this is the solution of the system;
\item Otherwise to go to step 1) until we find the solution.
\end{enumerate}

For clarity, we consider the numerical example of the remark \ref{r1} . If $n=4096$ and $h=6$, the step 1) given $2^{372}$  choices. The step 3) requires the xor of 61 words to compute a number of unknowns equal to 32 (64 if the machine works in 64 bits).

It is very likely that the values of these first 32 unknowns do not respect the \ref{iii}. Therefore we can estimate, although not accurately, the expected time to find the solution in $5,9\cdot 10^{113}$ machine cycles.

\subsection[Second strategy]{Second strategy}

The third strategy is consisted  in solving the system applying the XL relinearization algorithm by \cite{Joye}. In the case of the numerical example of remark \ref{r1} this algorithm produces a  system of $\bigg( \begin{matrix} 3968\\ 12\end{matrix}\bigg)$ linear equations, obtaining a computational complexity of $\bigg( \begin{matrix} 3968\\ 12\end{matrix}\bigg)^{2,37}=5,7\cdot 10^{81}$. So this strategy seems to be the best one.\\

We insisted that the particular structure of equations \ref{iii} can speed up above strategies, nevertheless the estimates seem to be quite reassuring.

\subsection[Third strategy]{Third strategy} 


Another possible strategy is to use linear equations to eliminate some of the variables $ x_i $ and derive the remaining $ x_i $ by using quadratic equations. In this case we have 8192 unknowns and 4224 (4096+128) linear equations. Therefore we cane delete (most) 4224 unknown. We remain with 3968 unknown and $64x63x64$ quadratic equations. We use the Brdet estimate. In this case $k \simeq 65$ and $D\simeq 7.6$. Let D=7 the estimate operations number to solve the system is $\bigg( \begin{matrix} 3968\\ 7\end{matrix}\bigg)^{2,37}\simeq 8.4 \cdot 10^{50}$ that is less of the previous strategies.

\subsection[Fourth strategy]{Fourth strategy} 


Another strategy to solve the system is probabilistic type. More precisely we can fix arbitrarily in each of the 128 blocks by 64 variables, 31 variables equal. Now we have a sufficient number of equations to determinate all variables. It is easy to controll if a variable is equal 1. The success probability is equal to  $\bigg(\frac{31}{64}\bigg)^{128}=1,5\cdot 10^{-37}=(6,6)\cdot 10^{36})^{-1}$.


If the computational cost of a single verification is $10^4$, we obtain a probability to solve the system equal to $\simeq 1-1/e$ with a  computational complexity of  $6,6 \cdot 10^{40}$.

%
%




\end{document}